# Cooperative O-H···π and C-H···O Hydrogen Bonding in Benzene-Methanol Solution: Strong Structures from Weak Interactions

Camilla Di Mino,[a] Andrew G. Seel,[b,] * Adam J. Clancy,[c] Andrea Sella,[c] Thomas F. Headen,[b] Támas Földes,[a] Edina Rosta,[a] Neal T. Skipper,[a,*]

(a) Department of Physics and Astronomy, University College London, London WC1E 6BT (UK),

(b) ISIS Neutron and Muon Source, Science and Technology Facilities Council, Rutherford Appleton Laboratory, Oxfordshire, OX11 0QX (UK)

(c) Department of Chemistry, University College London, 20 Gordon Street, London WC1H 0AJ (UK)

Email: andrew.seel@stfc.ac.uk; n.skipper@ucl.ac.uk.

**ABSTRACT**: Weak hydrogen bonds, such as O-H···π and C-H···O, are pivotal in a wide range of important natural and industrial processes including biochemical assembly, molecular recognition, and chemical selectivity. In this study we use neutron diffraction in conjunction with comprehensive H/D isotopic substitution to obtain a detailed spatial and orientational picture of the structure in benzene-methanol solution. This system provides us with a prototypical situation where the aromatic ring can act as an hydrogen bond acceptor (via the π electron density) and/or a hydrogen bond donor (via the CH groups), with the potential for cooperative effects. Our analysis places benzene at the centre of our frame-of-reference, and reveals for the first time that in solution the O-H···π interaction is highly localised and directional, the hydrogen atom being located directly above/below the ring centroid at a distance of 2.30 Å and with the hydroxyl bond axis normal to the aromatic plane. The tendency of methanol to form chain and cyclic motifs in the bulk liquid is manifest in a highly templated, symmetrical equatorial solvation structure; the methanol molecules surround the benzene so that the O-H bonds are coplanar with the aromatic ring while the oxygens interact with C-H groups through simultaneous bifurcated hydrogen bonds. By contrast, C-H···π interactions are relegated to the role of more distant spectators. The experimentally observed solvation therefore demonstrates that weak hydrogen bonding can give rise to strongly-ordered cooperative structural motifs also in the liquid phase.

## 1. Introduction

Hydrogen bonding systems are characterised by short X-H···Y interactions. Such bonds have traditionally been associated with X = N, O and have a strong electrostatic component arising from the highly dipolar nature of X-H bonds. However, in 1962 Jane Sutor reported the existence of short C-H···O contacts in the solid state, a discovery that opened the way for the exploration of a wide variety of weak hydrogen bonding interactions involving less electronegative elements acting as both donors and acceptors.[1,2] Beyond the traditional hydrogen bonding interaction, the highly polarised nature of the O-H bond allows the electron rich π system of an aromatic ring to act as a hydrogen bond acceptor. Such O/N/C-H···π interactions have been found in the solid state for both small molecule and biomolecule crystals and are now recognised as contributing significantly to molecular recognition and protein folding.[3–14]

Spectroscopic methods such as rotational microwave, IR, and NMR are traditionally used to identify such non-conventional interactions, including those involving aromatics and the -XH groups of molecules such as water, ammonia, and methanol. NMR in particular has produced advances in identifying weak hydrogen-bonding in crystalline phases via proton-mediated *J* couplings.[15,16] However, detecting such effects experimentally in the liquid phase presents formidable challenges due to rapid molecular reorganisations and competing interactions. In the case of $H_2O$ and $NH_3$ this is further compounded by their extremely low solubility in benzene, making a full characterisation of weak interactions very challenging.[13,17–22] Moreover, while these spectroscopic techniques point towards the existence of such bonding motifs, they do not allow a specific determination of the position of the hydrogens, nor do they reveal details of the effect of hydrogen bonding on solvation structures in the liquid state. These deficiencies can be addressed by neutron diffraction.

Previous research on bulk liquids via neutron diffraction has included studies of pure aromatic systems, such as benzene and toluene, and anisoles and its derivatives revealing the presence of π⋯π, C-H⋯π, C-H⋯O weak interactions. [23,24] In the last of these studies, the presence of an electron-donating methoxy group in anisole strongly affects the π cloud and the H-Me substitution doesn't provide the opportunity for O-H⋯π bonding.

To understand the underlying structural nature of weak hydrogen bonding interactions in the liquid state, we have therefore explored the archetypal and simplest fully miscible binary system: benzene in methanol. Indeed, *ab initio* studies on benzene-methanol molecular clusters propose three optimised minimum energy configurations exhibiting: O-H⋯π, C-H⋯π and C-H⋯O weak hydrogen bonding.[25] To answer the key question of which, if any, of these possible configurations are actually present in the liquid state, the current work exploits state-of-the-art neutron diffraction with selective isotopic substitution, supported by Monte Carlo molecular simulations refined against the experimental data. This approach enables us to determine the extent of structural order on a site-by-site basis, including the detailed relative position of hydroxyl groups around benzene. Our results elucidate the precise nature of two dominant non-covalent interactions between the OH group in-plane and out-of-plane from the π system.

**2. Experimental Methods**

Data have been acquired using the Near and InterMediate Range Order Diffractometer (NIMROD) at the ISIS Neutron and Muon Source (Didcot, UK) across a Q range of 0.05 Å$^{-1}$ – 50 Å$^{-1}$. [26] Absolute normalisation, instrument background, multiple and inelastic scattering have been corrected using standard procedures as implemented within the Gudrun package to obtain the total structure factors.[27] The extensive use of H/D isotopic substitution has permitted the acquisition of six isotopically distinct data sets. This approach provides strong constraints for structural refinement, with an atomistic model for the liquid system (50 benzene and 950 methanol molecules in a periodically repeated cubic box with sides 41.95 Å) having been refined simultaneously to each experimental dataset using the technique of Empirical Potential Structure Refinement (EPSR) (see Supporting Information section S3: EPSR method). [28] Specific site-centered Radial Distribution Functions (RDFs), Angular Radial Distribution Functions (ARDFs) and Spatial Density Functions (SDFs) were extracted from our experimentally-derived structural models using the DLputils and Aten packages (see SI for function definitions). [29,30]

## 3. Results and Discussion

The experimental neutron diffraction structure factors, $F(Q)$, and EPSR refinements are presented for the six isotopically varied systems alongside their total pair distribution functions, $G(r)$, in Figure 1. Excellent agreement has been achieved between model and experimental data for each sample dataset; small discrepancies at low-$Q$ are attributed to a residual presence of inelastic and multiple scattering events.

The existence of O-H⋯π hydrogen bonding is clearly evident from Figure 2 reporting the benzene centre-of-mass (CoM) - methanol partial distribution functions extracted from the refined structural model for the H, O and C sites of the methanol.[31] The closest approach of the hydrogen to the benzene CoM arises at 2.30 Å as shown in the $g_{CoM-H}$ (r). We conclude that the methanol protons must approach from directly above and below the plane of the benzene ring, as the observed CoM - H distance is too short to allow in-plane approach. The $g_{CoM-O}$ (r) demonstrates that the closest oxygen atom to the benzene ring is found at 3.25 Å. This is 0.95 Å further than the CoM-H distance and therefore precisely within the window of H-O intra-molecular covalent bonding length, implying that the ring CoM···H-O contact is linear.[32] The average coordination number of these hydrogen atoms within a distance of 3.5 Å from the CoM of the benzene is found to be $N_{CoM-H}$ = 0.59 ± 0.05 (see SI section S5: Coordination Numbers). Since benzene has two π-faces, it can accommodate up two ring CoM···H-O contacts, and there is therefore the possibility for each molecule to accept 0, 1 or 2 hydrogen bonds with H(-O) donors.[33] To determine the relative likelihoods of these three possible scenarios in solution, we have interrogated the EPSR configurations of our system molecule-by-molecule. This analysis indicates that the probabilities of 0, 1 and 2 hydrogen bonds are 0.51, 0.41 and 0.08 respectively. The first of these values highlights that such CoM···H-O interactions are indeed weak in the liquid state. This conclusion is further supported by the overall probability distribution, which is consistent with the two π-faces of each benzene acting independently in this context (there is no probabilistic preference for 1 rather than 2 hydrogen bonds). Our results are therefore consistent with an O-H···π interaction of electrostatic nature, where the donor interacts with the acceptor centroid linearly. [33]

The second peaks of the hydrogen and oxygen partial distribution functions appear equidistant at 4.9 Å. These distances correspond to both the primary solvation shell in the plane of the benzene ring and the second solvation shells above and below the ring. Importantly the $g_{CoM-C}$ (r) indicates that the methyl groups are found at further distances from the benzene molecule than either hydrogen or oxygen atoms, indicating a strong preference for solvation of the benzene by the hydroxyl groups.

The local structure of methanol around benzene can therefore be partitioned into two distinct environments: in-plane and out-of-plane to the ring. To understand the spatial and orientational arrangement of methanol in the environments, we can examine the ARDFs, the partial distribution functions dependent on the relative orientation of the two molecules, and the SDFs, the three-dimensional density distribution between two defined species. In particular, in Figure 3 $a_1$, $b_1$, $c_1$ we can determine the relative orientation of the $C_6$ axis of the benzene and the O-H/C-O vectors. Figure 3 $a_1$ shows two sharp peaks at 0˚ and 180˚ at approximately 3 Å distance, confirming that, in absence of steric hindrance from either internal or external components, the O-H approaches the π system directly along the principal axis. The linearity of the interaction is further confirmed in the SDF of Figure 3 $a_2$.

To characterise the interactions in the plane of the ring, we examine the ARDF for the orientation of the O-H vector of the methanol relative to the molecular plane of benzene. This benzene to methanol O-H ARDF is presented in Figure 3 $b_1$, and shows a peak centred at 90˚, allowing both methanol O and H sites to remain in the benzene plane and at similar distances to the benzene CoM. The refined structure for the solvation of benzene is presented in Figure 3 $b_2$. The oxygen atom acts here as a bifurcated H-bond acceptor, bonding simultaneously with two hydrogen atoms on the benzene at a distance of 2.9 Å (Figure S2), precluding the presence of a traditional linear

C-H⋯O weak hydrogen bond between C-H and O. [3] As a result, the observed configuration of the first benzene in-plane solvation shell is an equatorial 'belt like' structure around the ring, consistent with the known tendency of methanol to form long chains both in bulk and in the presence of benzene. [34,35] This belt-like effect arises from cooperative O-H⋯π, C-H⋯O interactions, first identified in crystalline phases. We note that this stabilising behaviour is not predicted by classical molecular dynamics simulations, nor Monte Carlo simulations absent the refinement from neutron scattering profiles (Figure S8, S9). We have now shown that such interactions play an important role in solution structures. [36,37]

Figure 3 $c_{1,2}$ shows the most likely position of the methyl groups surrounding the benzene molecule. The ARDFs (Figure 3 $c_1$) show that the relative angle between the $C_6$ axis of the benzene ring and the C-O vector is at 60° – 120° which translates to a ring-shaped spatial density above and below the benzene plane in the SDFs of Figure 3 $c_2$. Furthermore, the methyl groups are found around the ring in a complementary position to the O and H sites, in agreement with the belt-like structure, reinforcing the conclusion that solvation of benzene takes place via the hydroxyl groups rather than via methyl-groups. This is clearly an indication of how much more significant dipolar interactions are compared to dispersion in this system.

Comparing our experimentally derived solvation structures with those proposed by earlier *ab initio* for benzene-methanol clusters, we note that the observed O-H⋯π is directed towards the ring centre, rather than the ring carbons. [25] It is also revealed that, despite not being the minimum energy configuration, the bifurcated coplanar structure is not disfavoured. [3,38] It is worth mentioning that *ab initio* studies on benzene-methanol clusters include the presence of another favourable configuration, which involves a C-H⋯π weak hydrogen bonding between the methyl hydrogens and the benzene ring.[25] This interaction is absent from the *g(r)* and ARDFs of Figure 2 and 3 as it is weaker than the competing O-H⋯π and C-H⋯O interactions present in the current system. In addition, the classic benzene-benzene interactions seen in the bulk liquids, namely for benzene face-to-face stacking at shorter distances and Y-stacking at longer distances, are maintained in the mixture (Figure S2). [24,39]

## 4. Conclusion

In summary, neutron diffraction in conjunction with H/D isotopic labelling has been used to reveal the rich hydrogen-bonded structure of benzene-methanol solutions. Our analysis is supported by molecular modelling refined against the experimental data. Our study provides detailed insights into the spatial and orientational configurations that occur in solution. We propose a new solvation structure for benzene, in which a highly directional O-H⋯π bond sits directly above/below the benzene CoM and aligned normal to the aromatic plane. This motif is strongly localised relative to the ring centre, with well-defined H and O distances of 2.30 Å and 3.25 Å respectively. In addition, we find that methanol forms an equatorial belt around benzene, in which C-H⋯O interactions are bifurcated and the solvent aligns so that O-H bonds are parallel to the aromatic plane. We therefore conclude that, even in the liquid state, weak interactions can give rise to well-defined intermolecular structural motifs.

**ASSOCIATED CONTENT**

**Supporting Information.**

Complementary Neutron Diffraction Theory; Experimental Method including neutron weights; The Empirical Potential Structure Refinement (EPSR) Method including Lennard-Jones seed parameters; Angular Radial Distribution Functions and Spatial Distribution Functions including definition of distribution functions and axes and supplementary partial radial distribution functions; Coordination Numbers; Molecular Dynamics (MD) and Monte Carlo (MC) Simulations without empirical potential refinement.

# AUTHOR INFORMATION

**Corresponding Authors**

Andrew G. Seel: andrew.seel@stfc.ac.uk

Neal T. Skipper: n.skipper@ucl.ac.uk

# ACKNOWLEDGMENT

The authors acknowledge the UK Science and Technology Facilities Council (STFC) for NIMROD beamtime allocation (10.5286/ISIS.E.RB1510644) and the use of SCARF computational facility for EPSR simulations. AJC thanks the Ramsay Memorial Fellowship Trust and the Royal Society University Research Fellowship for funding. Engineering and Physical Sciences Research Council (EPSRC, grant EP/R513143/1) is acknowledged for support of a PhD studentship for CDM. The authors thank Tristan G. A. Youngs, Christopher A. Howard and Daniele Paoloni for useful discussions.

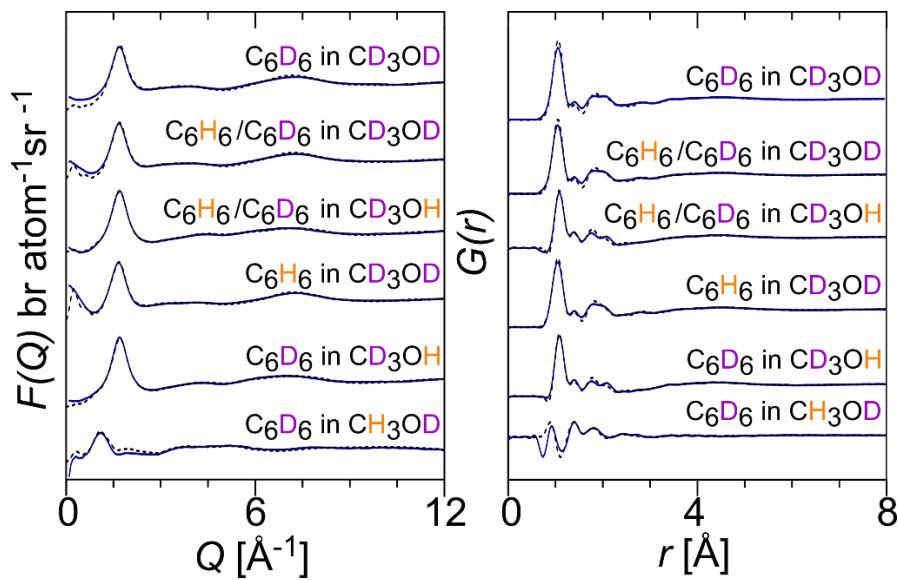

**Figure 1.** Experimental (solid) and modelled (dashed) neutron diffraction partial structure factors (left) and total radial distribution functions (right) for the six isotopically distinct samples. Note how the use of isotope H/D labelling allows us to change the relative contributions from the different atomic sites (see Table S1 for full neutron weightings), thereby enabling us to resolve subtle structural effects in the benzene-methanol solution.

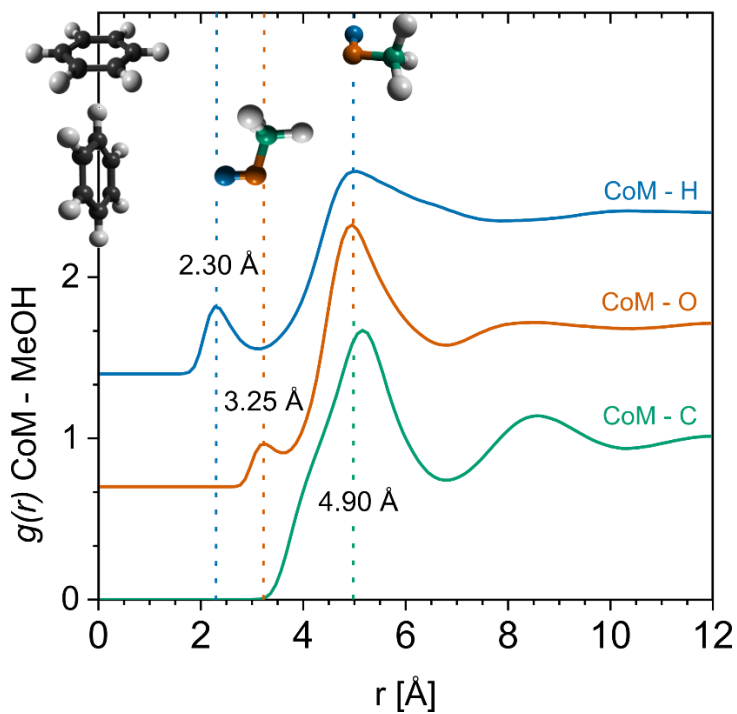

**Figure 2.** Benzene CoM-MeOH Partial Distribution Functions: from top to bottom CoM-H(O) blue, CoM-O(H) orange, CoM-C green. A schematic representation of the primary interactions is shown, highlighting the comparable distance between the H/O *g(r)s* peaks and the intra-molecular O-H bonding length and the in-plane position of H/O atoms at further distances.

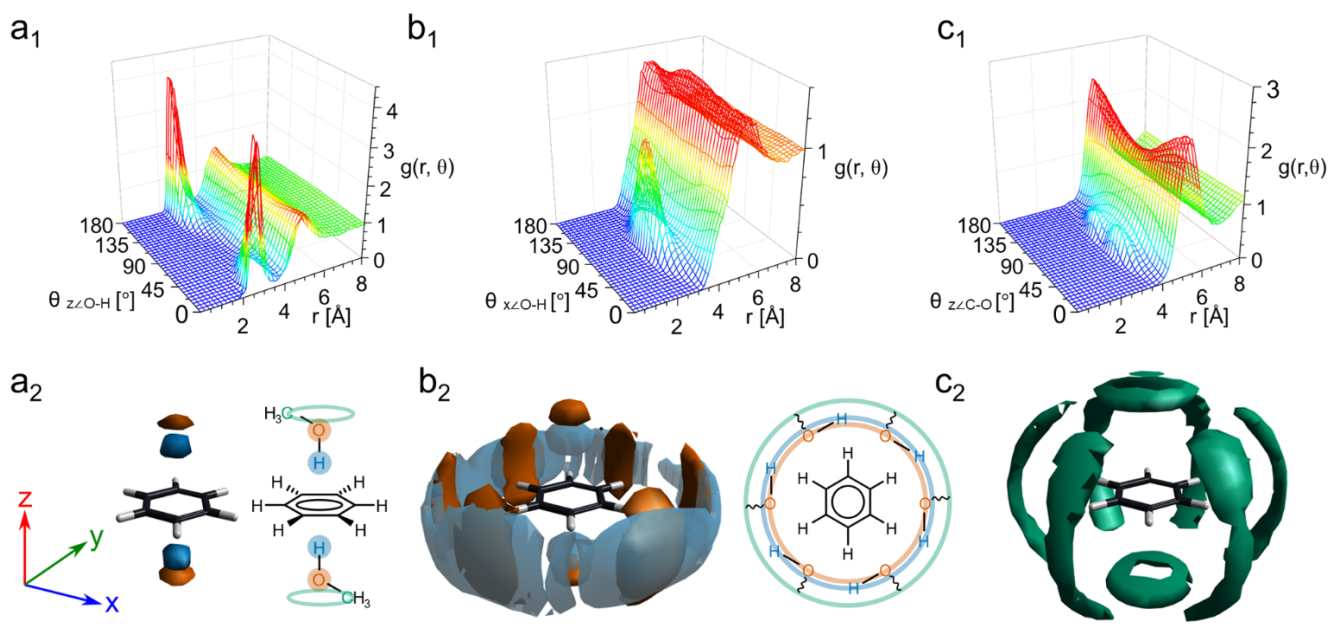

**Figure 3.** $a_1$, $b_1$ and $c_1$ Angular Radial Distribution Functions (ARDFs) for benzene ring centre - OH methanol, $g(r, \vartheta)$ where r is the distance between the sites and $\vartheta$ is the angle between the normal to the ring plane (z axis) and the O-H ($a_1$, $b_1$) and C-O axis $c_1$). ($a_2$, $b_2$, $c_2$) Spatial Density Functions (SDFs) showing the most likely positions for molecules in the first coordination shell: $a_2$ shows the 30% most likely position methanol hydroxyl H (dark blue) and methanol O (dark orange) up to 3.1 Å and 3.7 Å, respectively from the benzene CoM; $b_2$ shows the 5% most likely position of methanol H and methanol O up to 6.8 Å from the benzene CoM; $c_2$ shows the 10% most likely position of methanol C up to 6.8 Å from the benzene CoM. See Figure S3 for comprehensive axes definition.